\documentclass{pasj00}
\def\Umlaut#1{\"{#1}}

\def\commenta{$^*$}

\def\inpress{in press}
\def\astroph#1{ (astro-ph/#1)}

\DeclareAbbreviation\AAHam{Astron. Abh. Hamburg. Sternw.}
\DeclareAbbreviation\AARv{Astron. Astrophys. Rev.}
\DeclareAbbreviation\an{Astron. Nachr.}
\DeclareAbbreviation\AcA{Acta Astron.}
\DeclareAbbreviation\Afz{Astrofizika}
\DeclareAbbreviation\AnTok{Tokyo Astron. Obs. Annals, Sec. Ser.}
\DeclareAbbreviation\Ap{Astrophysics}
\DeclareAbbreviation\ARep{Astron. Rep.}
\DeclareAbbreviation\ATel{Astronomer's Telegram}
\DeclareAbbreviation\ATsir{Astron. Tsirk.}
\DeclareAbbreviation\AcApS{Acta Astrophys. Sinica}
\DeclareAbbreviation\AstL{Astron. Letters}
\DeclareAbbreviation\BaltA{Baltic Astron.}
\DeclareAbbreviation\BASI{Bull. Astron. Soc. India}
\DeclareAbbreviation\BeSN{Be Star Newsletter}
\DeclareAbbreviation\GCN{GCN}
\DeclareAbbreviation\ibvs{Inf. Bull. Variable Stars}
\DeclareAbbreviation\JAD{J. Astron. Data}
\DeclareAbbreviation\JAVSO{J. American Assoc. Variable Star Obs.}
\DeclareAbbreviation\JBAA{J. British Astron. Assoc.}
\DeclareAbbreviation\LowOB{Lowell Obs. Bull.}
\DeclareAbbreviation\MitVS{Mitteil. Ver\"{a}nderl. Sterne}
\DeclareAbbreviation\MmSAI{Mem. Soc. Astron. Ita.}
\DeclareAbbreviation\Msngr{Messenger}
\DeclareAbbreviation\NewA{New Astron.}
\DeclareAbbreviation\NewAR{New Astron. Rev.}
\DeclareAbbreviation\OAP{Odessa Astron. Publ.}
\DeclareAbbreviation\Obs{Observatory}
\DeclareAbbreviation\PASA{Publ. Astron. Soc. Australia}
\DeclareAbbreviation\PAZh{Pis'ma AZh}
\DeclareAbbreviation\PhR{Phys. Rep.}
\DeclareAbbreviation\PVSS{Publ. Variable Stars Sect. R. Astron. Soc. New Zealand}
\DeclareAbbreviation\PZ{Perem. Zvezdy}
\DeclareAbbreviation\PZP{Perem. Zvezdy Pril.}
\DeclareAbbreviation\QJRAS{QJRAS}
\DeclareAbbreviation\RMxAA{Rev. Mexicana Astron. Astrof.}
\DeclareAbbreviation\RvMA{Reviews of Modern Astron.}
\DeclareAbbreviation\Sci{Science}
\DeclareAbbreviation\SvA{Soviet Astronomy}
\DeclareAbbreviation\SvAL{Soviet Astronomy Letters}
\DeclareAbbreviation\VeSon{Ver\"{o}ff. Sternw. Sonneberg}
\DeclareAbbreviation\VSOLJBul{VSOLJ Variable Star Bull.}
\DeclareAbbreviation\yCat{VizieR Online Data Catalog}
\DeclareAbbreviation\ZA{Z. Astrophys.}

\def\ASPConf#1#2{ASP Conf. Ser. #1, #2}

\def\PublisherASP{San Francisco: ASP}
\def\PublisherReidel{Dordrecht: D. Reidel Publishing Company}

\begin{document}

\SetRunningHead{T. Kato}{1993 Superoutburst of LL Andromedae}

\Received{}%{yyyy/mm/dd}
\Accepted{}%{yyyy/mm/dd}

\title{1993 Superoutburst of LL Andromedae}

\author{Taichi \textsc{Kato}}
\affil{Department of Astronomy, Kyoto University,
       Sakyo-ku, Kyoto 606-8502}
\email{tkato@kusastro.kyoto-u.ac.jp}

%%% end:list of authors

\KeyWords{accretion, accretion disks
          --- stars: dwarf novae
          --- stars: individual (LL Andromedae)
          --- stars: novae, cataclysmic variables
}

\maketitle

\begin{abstract}
  We present time-resolved CCD photometry of LL And during its 1993
outburst.  The observation revealed the presence of superhumps with a
period of 0.05697(3) d.  This period is one of the smallest among the
hydrogen-rich dwarf novae.  Although LL And has been proposed to be
a WZ Sge-type dwarf nova based on its low outburst frequency, our new
analysis indicates that the outburst amplitude ($\sim$ 5 mag) and
outburst duration (9$\pm$2 d) are much smaller and shorter than in
typical WZ Sge-type dwarf novae.
We suspect that the unusual outburst properties of LL And
might be explained by assuming a ``leaky disk" in quiescence, which
was originally proposed to explain the prototypical
WZ Sge-type outbursts.
By combination with the recent suggestion of the orbital period,
the fractional superhump excess is found to be 3.5(1) \%, which is
unusually large for this short-period system.  LL And may be an object
filling the gap in the evolutionary track, which has recently been
proposed to explain the unusual ultracompact binaries with an evolved
mass donor.
\end{abstract}

\section{Introduction}\label{sec:intro}

   WZ Sge-type dwarf novae are a class of SU UMa-type
dwarf novae [for recent summaries of dwarf novae and SU UMa-type dwarf
novae, see \citet{osa96review} and \citet{war95suuma}, respectively],
characterized by a long ($\sim$ 10 yr) outburst
recurrence time and a large ($\sim$ 8 mag) outburst amplitude
(cf. \cite{bai79wzsge}; \cite{dow81wzsge}; \cite{pat81wzsge};
\cite{odo91wzsge}; \cite{kat01hvvir}).  WZ Sge-type dwarf novae
are considered to be systems close to the terminal evolution
of cataclysmic variables (CVs).  Since the expected mass of the
mass-donor secondary stars in such systems is close to the lower
limit of normal low-mass stars, WZ Sge-type dwarf novae are recently
regarded as promising candidates for binaries containing
brown dwarfs (\cite{how97periodminimum}; \cite{pol98TOAD};
\cite{cia98CVIR}; \cite{pat01SH}; \cite{how01llandeferi};
\cite{men02CVBD}; \cite{lit03CVBD}).  LL And, being known as
a hydrogen-rich dwarf nova with one of the shortest periods, has
been nominated as a promising candidate harboring a brown dwarf
secondary \citep{how01llandeferi}.

   LL And is an eruptive object discovered in 1979 \citep{wil79lland}.
Since the only approximate position was announced at the time of the
discovery, we examined the Palomar Observatory Sky Survey (POSS) I
prints and identified a blue object close to the reported position
(T. Kato, 1990, unpublished).  This supposed identification, which was
later confirmed by the detection of a new outburst in 1993,
of a relatively bright ($\sim$ 19 mag) quiescent counterpart naturally
suggested the dwarf nova-type classification.
This information was quickly relayed
to observers through the international alert networks
(e.g. VSNET: \cite{VSNET}),
and the object has been continuously monitored since then.
The long-awaited next outburst finally occurred in 1993.

\section{Observation}

   The CCD photometric observation was performed using a CCD camera
(Thomson TH~7882, 576 $\times$ 384 pixels, on-chip 2 $\times$ 2 binning
adopted) attached to the Cassegrain focus of the 60-cm reflector
(focal length = 4.8 m) at Ouda Station, Kyoto University \citep{Ouda}.
The frames were first corrected for standard de-biasing
and flat fielding, and were then processed by a microcomputer-based
photometry package developed by the author.  The log of observations
is listed in table \ref{tab:log}.

\begin{table*}
\caption{Journal of CCD photometry.}\label{tab:log}
\begin{center}
\begin{tabular}{ccrcccrcc}
\hline\hline
\multicolumn{3}{c}{Date}& Start--End\commenta & Filter & Exp(s) & $N$
        & Mean mag & Error \\
\hline
1990 & December & 30 & 48255.992--48255.993 & $V$ & 60--120 & 2 & [18.0 & $\cdots$ \\
1991 & July & 19 & 48457.212--48457.217 & $I_{\rm c}$ & 120 & 3 & [18.0 & $\cdots$ \\
     & August & 21 & 48490.208--48490.215 & $I_{\rm c}$ & 180 & 4 & 18.8 & 0.5 \\
     &        & 31 & 48500.137--48500.138 & $I_{\rm c}$ & 60--90 & 2 & [18.0 & $\cdots$ \\
1993 & December &  9 & 49330.863--49331.096 & $V$ & 120--240 &  96 & 14.54 & 0.01 \\
     &          & 11 & 49332.904--49333.060 & $V$ & 120      &  69 & 14.62 & 0.01 \\
     &          & 12 & 49333.877--49334.116 & $V$ & 120      & 147 & 14.65 & 0.01 \\
     &          & 19 & 49340.948--49341.063 & $V$ & 120--240 &  25 & 17.63 & 0.04 \\
1994 & January  &  1 & 49353.882--49353.964 & $V$ & 60--480  &  20 & 18.07 & 0.05 \\
     &          &  2 & 49354.905--49355.001 & $V$ & 60--360  &  22 & 18.33 & 0.04 \\
\hline
 \multicolumn{9}{l}{\commenta JD$-$2400000.} \\
\end{tabular}
\end{center}
\end{table*}

   The observations used GSC 1741.858 ($V$ = 13.87) as the comparison star,
and GSC 1741.301 ($V$ = 14.58) and GSC 1741.441 ($V$ = 14.62) as the
check stars.  The magnitudes of the comparison stars were taken
from \citet{hen97sequence}.  The VSNET chart on this magnitude system
(showing LL And in outburst) is available at
$\langle$ftp://vsnet.kusastro.kyoto-u.ac.jp/pub/vsnet/charts/LL\_And.ps$\rangle$.  The GSC magnitudes, which were used at the time of the 1993 outburst,
are confirmed to be $\sim$ 0.4 mag too bright, which explains the
discrepancy between the present CCD $V$ measurements and the visual
observations published on the 1993 outburst occasion \citep{how94lland}.
We thereby added a correction of 0.4 mag to the visual observations
described in \citep{how94lland}.  We also obtained a few snapshot
quiescent observations which are also listed in table \ref{tab:log}.

   Heliocentric corrections to the observed times were applied before the
following period analysis.

\section{Results}

\subsection{Outburst Light Curve}

\begin{figure}
  \begin{center}
%    \FigureFile(88mm,60mm){lc.eps}
    \FigureFile(88mm,60mm){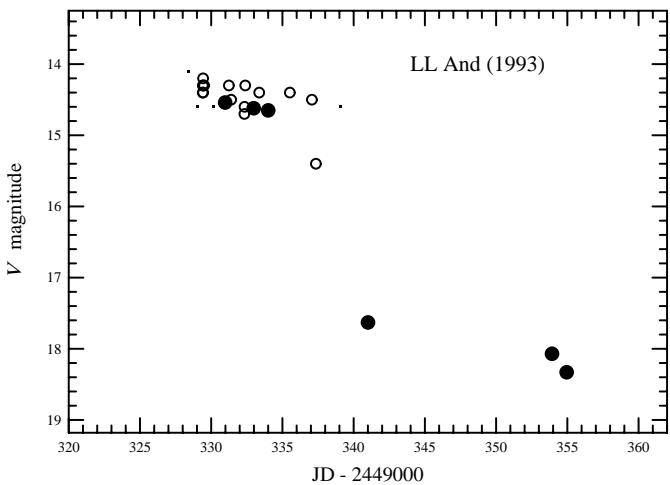}
  \end{center}
  \caption{Light curve of the 1993 outburst of LL And.
  The large filled circles and open circles represent our CCD $V$-band
  and visual observations (corrected for the 0.4 mag zero-point error in
  the comparison stars), respectively.  The small dots represent upper
  limit observations.
  }
  \label{fig:lc}
\end{figure}

   The resultant outburst light curve is presented in figure \ref{fig:lc}.
The object was first detected in outburst by T. Vanmunster on 1993
December 7.906 UT.  The object was not detected in outburst 0.34 d
(M. Moriyama), 1.03 d (P. Schmeer), 18.00 d (G. Poyner) before this
detection, respectively.  It was thus less likely that the object
underwent a missed brighter maximum before Vanmunster's initial
detection.  Following a slow decline at least until December 15,
the object entered the rapid decline stage sometime before December 19.
The most likely duration of the outburst was 9$\pm$2 d.

\subsection{Superhumps}

\begin{figure}
  \begin{center}
%    \FigureFile(88mm,120mm){shlc.eps}
    \FigureFile(88mm,120mm){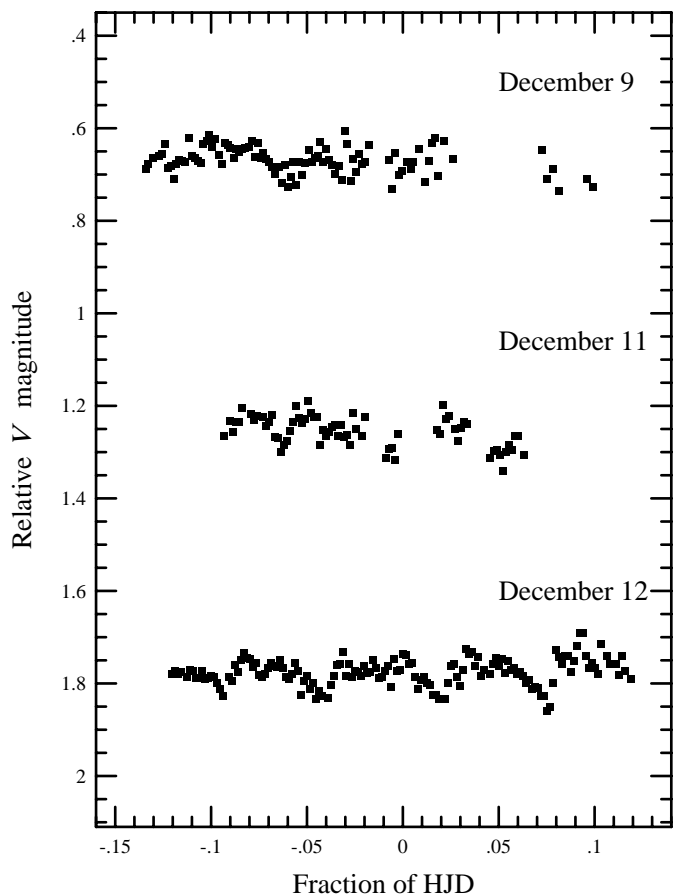}
  \end{center}
  \caption{Superhumps in LL And.  The errors of individual observations
  are smaller than the size of the marks.}
  \label{fig:shlc}
\end{figure}

   On December 9, 11, and 12, our time-resolved CCD photometry revealed
the presence of superhumps.  This information, relayed through the
alert network system \citep{VSNET}, was announced by \citet{how94lland}.
The nightly light curves of the superhumps are shown in
figure \ref{fig:shlc}.

   After removing the trend of linear decline (0.64 mag d$^{-1}$),
the December 9--12 observations were analyzed using
the Phase Dispersion Minimization (PDM: \cite{PDM}).  The resultant
period is 0.05697(3) d.  The statistical significance of the period
is better than 99.5\%.  The selection of the correct alias was confirmed
by independent analyses of continuous nightly observations.
The error of the period was estimated using the Lafler--Kinman class of
methods, as applied by \citet{fer89error}.  This period supersedes
the older preliminary values cited in \citet{how94lland} and
\citet{kat01hvvir}.

\begin{figure}
  \begin{center}
%    \FigureFile(88mm,60mm){pdm.eps}
    \FigureFile(88mm,60mm){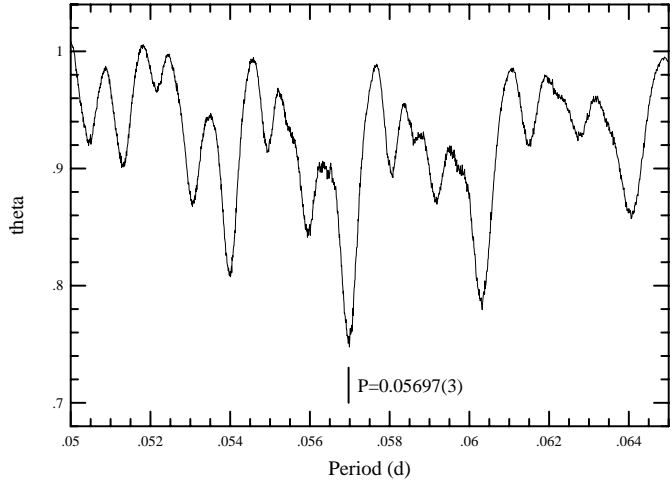}
  \end{center}
  \caption{Period analysis of superhumps in LL And.
  }
  \label{fig:pdm}
\end{figure}

\begin{figure}
  \begin{center}
%    \FigureFile(88mm,60mm){ph.eps}
    \FigureFile(88mm,60mm){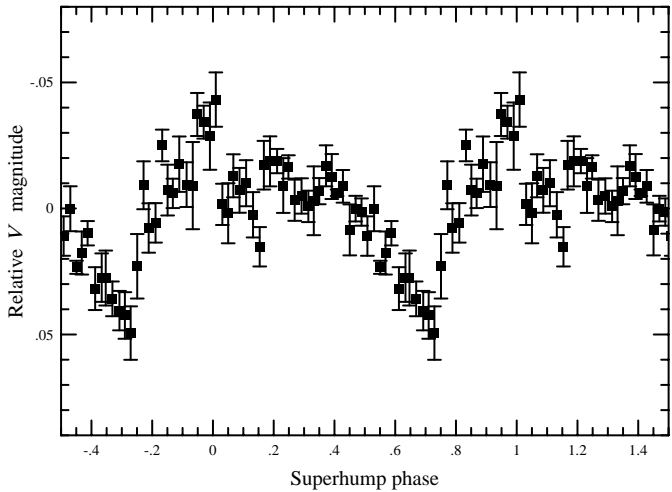}
  \end{center}
  \caption{Averaged superhump profile of LL And.
  }
  \label{fig:ph}
\end{figure}

   Figure \ref{fig:ph} shows the phase-averaged superhump profile of
LL And folded with the best period of 0.05697 d.  Although the profile
is rather complex (with a secondary maximum following the primary
maximum), the overall appearance with a faster rise and and a slower
decline is characteristic of SU UMa-type superhumps
(\cite{vog80suumastars}; \cite{war85suuma}).

\section{Discussion}

\subsection{Outburst Properties}\label{sec:outprop}

   \citet{how94lland}, \citet{how96lland} suggested, from the available
material at these times, that LL And belongs to a class of dwarf novae
with large outburst amplitudes.  This identification, however, becomes
dubious upon closer examination of the present material.

   Firstly, \citet{how94lland} used the maximum outburst magnitude of
$m_{\rm vis}$ = 13.8, which is clearly an overestimated caused by
an incorrect zero point.  The present observation, calibrated on the
modern $V$ scale, suggests a much fainter outburst maximum of
$V$ = 14.3--14.5.  The bright magnitude quoted by \citet{wil79lland}
needs to be treated with special caution, because the observation
probably used blue-sensitive plates (hence would not adequately
represent visual magnitudes), and because the published magnitudes
were very likely only preliminary measurements with probable errors
of $\sim$1 mag.

   Secondly, the quiescent magnitude in \citet{how94lland},
\citet{how96lland} was likely underestimated.  The object is already
readily recognized on paper reproduction of POSS I red and blue prints
(section \ref{sec:intro}), which suggests a significantly brighter
magnitude than $V \sim$ 20.  The modern magnitude estimates
(USNO B1.0: \cite{USNOB10}) give red and blue magnitudes of
19.26 and 19.59--19.78, respectively.  These measurements are in line
with the author's estimate on POSS I paper prints.  The USNO B1.0
magnitude correspond to $V$ = 19.4.

   The inferred outburst amplitude from these new estimates is $\sim$5.0
mag, which is no longer an exceptionally large value for SU UMa-type
dwarf novae (e.g. \cite{nog97sxlmi}).

   The outburst frequency looks like to be small.  The only recorded
outbursts up to now were in 1979 September \citep{wil79lland} and
in 1993 December (this work).  In spite of intensive monitoring mainly
by the VSNET \citep{VSNET} members, no definite outburst has been
recorded up to 2003.  Even considering the unavoidable seasonal
observational gaps, the detected outbursts are much less frequent than
in most dwarf novae, and may be comparable to those of the WZ Sge-type
dwarf novae.

\subsection{Fractional Superhump Excess}\label{sec:shexcess}

   Very recently, \citet{pat03suumas} reported the detection of photometric
periodicity of 0.055053(6) d.  Assuming that this periodicity represents
the orbital period ($P_{\rm orb}$),\footnote{
  One should be, however, careful in interpreting quiescent periodicity.
  The well-known WZ Sge-type object AL Com showed seemingly coherent
  variations whose period is clearly
  different from the supposed orbital period \citep{abb92alcomcperi}.
  We assume $P$ = 0.055053(6) d to likely represent $P_{\rm orb}$ because
  of its proximity to what would be expected from the superhump period
  using the known relation \citep{StolzSchoembs}.
} the fractional superhump excess $\epsilon=P_{\rm SH}/P_{\rm orb}-1$
amounts to 3.5(1) \%.  This value is exceptionally large for an SU UMa-type
system with $P_{\rm SH}$ = 0.05697 d (cf. \cite{pat01SH}).
This conclusion seems to further support the presence of a rather massive
secondary star \citep{pat03suumas}, and is likely incompatible with
the earlier claim of a brown-dwarf secondary
\citep{how01llandeferi}.\footnote{
  This claim was later questioned by the author themselves
  \citep{how02llandefpegHST}, and was more convincinglu refuted by
  \citet{lit03CVBD}.
}

\subsection{Comparison with Other Unusual Dwarf Novae}

   As shown in subsection \ref{sec:outprop}, the outburst cycle
length of LL And is likely comparable to rarely outbursting WZ Sge-type
dwarf novae.  The short superhump period (0.05697 d) is also comparable
to those of WZ Sge-type dwarf novae \citep{kat01hvvir}.  The object,
however, shows remarkable difference from typical WZ Sge-type
dwarf novae in its small ($\sim$ 5 mag) outburst amplitude
(compared to $\sim$ 8 mag for WZ Sge-type dwarf novae),
short (9$\pm$2 d) duration of the superoutburst
(compared to $>$20 d for WZ Sge-type dwarf novae,
cf. \cite{ish02wzsgeletter}; \cite{pat96alcom}; \cite{nog97alcom};
\cite{kat02v592her}).

   The combination of long outburst cycle length, low outburst amplitude,
and short duration of a superoutburst resembles that of an unusual
SU UMa-type dwarf nova GO Com (A. Imada et al., in preparation).
In GO Com, the small scale of the recorded outburst,
in spite of the long preceding quiescence,
is interpreted as the possible consequence of the extraction of
disk mass (e.g. via evaporation) during quiescence (A. Imada et al.,
in preparation).  This scenario was initially proposed by
\citet{las95wzsge} to explain the long outburst intervals in
systems resembling WZ Sge-type stars, but now looks more
applicable to systems such as GO Com and LL And
[see also discussions by \citet{osa95wzsge};
\citet{osa98suumareviewwzsge} on the difficulty
of reproducing WZ Sge-like outbursts with a ``leaky" accretion disk,
as in \citet{las95wzsge}].

\begin{figure*}
  \begin{center}
%    \FigureFile(160mm,80mm){excess.eps}
    \FigureFile(160mm,80mm){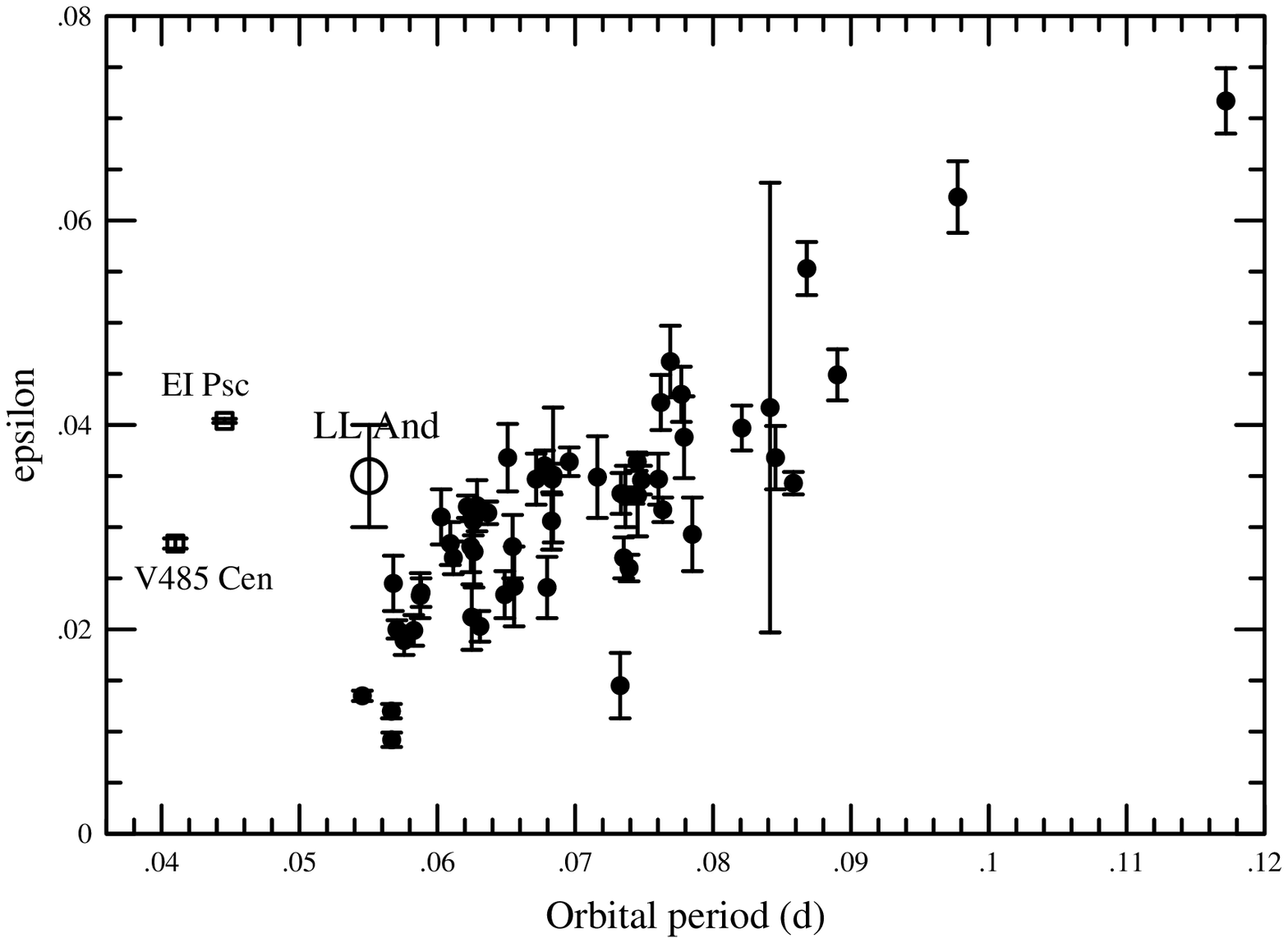}
  \end{center}
  \caption{Relation between orbital period ($P_{\rm orb}$) and fractional
  superhump period excess ($\epsilon$).  The basic data were mainly taken
  from \citet{pat98evolution}; \citet{pat01SH};
  \cite{tho02gwlibv844herdiuma}; \cite{pat03suumas}, supplemented and
  refined with \citet{uem02j2329letter}; \citet{kat02cccnc};
  \citet{kat03hodel}.  The small filled and open squares represent
  ordinary SU UMa-type dwarf novae, and unusual hydrogen-rich ultracompact
  binaries (EI Psc and V485 Cen), respectively.  The location of LL And
  is marked with a open circle.
  }
  \label{fig:excess}
\end{figure*}

   By adopting the large fractional superhump excess
(subsection \ref{sec:shexcess}), the secondary star of LL And is
likely slightly too massive for this period (figure \ref{fig:excess}).
We know at least two
well-established examples of such short-period dwarf novae with
unusually massive or luminous secondaries
(EI Psc: \cite{uem02j2329letter}; \cite{tho02j2329};
\cite{ski02j2329} and QZ Ser: \cite{tho02qzser}).  Both objects have
low outburst frequencies than would be expected from their binary
parameters.  \citet{uem02j2329letter} and \citet{tho02j2329} suggested
that EI Psc may be the first identified object following the hypothetical
evolutionary track \citep{pod03amcvn} containing an mass donor having
an evolved core.
LL And may be an object filling the evolutionary missing link
between QZ Ser and EI Psc (see also figure 4 in \cite{tho02qzser}),
and finally to the double-degenerate
AM CVn stars (\cite{war95amcvn}; \cite{sol95amcvnreview}).
Further spectroscopic determination of the orbital parameters
is encouraged.

\vskip 3mm

   The author is grateful to T. Vanmunster for promptly notifying us
of the rare outburst of LL And.  We are also grateful to a number of
observers who have been reporting their observations to the VSNET,
and to Dr. T. Takata for helping the observation.
This work is partly supported by a grant-in-aid (13640239, 15037205)
from the Japanese Ministry of Education, Culture, Sports, Science and
Technology.
This research has made use of the Digitized Sky Survey producted by STScI,
and the VizieR catalogue access tool.


\begin{thebibliography}{}

\bibitem[Abbott et~al.(1992)]{abb92alcomcperi}
  Abbott, T. M.~C., Robinson, E.~L., Hill, G.~J., \& Haswell, C.~A.\ 1992,
  \apj, 399, 680

\bibitem[Bailey(1979)]{bai79wzsge}
  Bailey, J.\ 1979, \mnras, 189, 41P

\bibitem[Ciardi et~al.(1998)]{cia98CVIR}
  Ciardi, D.~R., Howell, S.~B., Hauschildt, P.~H., \& Allard, F.\ 1998, \apj,
  504, 450

\bibitem[Downes, Margon(1981)]{dow81wzsge}
  Downes, R.~A., \& Margon, B.\ 1981, \mnras, 197, 35P

\bibitem[Fernie(1989)]{fer89error}
  Fernie, J.~D.\ 1989, \pasp, 101, 225

\bibitem[Henden, Honeycutt(1997)]{hen97sequence}
  Henden, A.~A., \& Honeycutt, R.~K.\ 1997, \pasp, 109, 441

\bibitem[Howell, Ciardi(2001)]{how01llandeferi}
  Howell, S.~B., \& Ciardi, D.~R.\ 2001, \apjl, 550, L57

\bibitem[Howell et~al.(2002)]{how02llandefpegHST}
  Howell, S.~B., G\Umlaut{a}nsicke, B.~T., Szkody, P., \& Sion, E.~M.\ 2002,
  \apj, 575, 419

\bibitem[Howell, Hurst(1994)]{how94lland}
  Howell, S.~B., \& Hurst, G.~M.\ 1994, \ibvs, 4043

\bibitem[Howell, Hurst(1996)]{how96lland}
  Howell, S.~B., \& Hurst, G.~M.\ 1996, \JBAA, 106, 29

\bibitem[Howell et~al.(1997)]{how97periodminimum}
  Howell, S.~B., Rappaport, S., \& Politano, M.\ 1997, \mnras, 287, 929

\bibitem[Ishioka et~al.(2002)]{ish02wzsgeletter}
  Ishioka, R., {et~al.}\ 2002, \aap, 381, L41

\bibitem[Kato et~al.(2003a)]{kat03hodel}
  Kato, T., Nogami, D., Moilanen, M., \& Yamaoka, H.\ 2003a, \pasj,
  \inpress\astroph{0307064}

\bibitem[Kato et~al.(2001)]{kat01hvvir}
  Kato, T., Sekine, Y., \& Hirata, R.\ 2001, \pasj, 53, 1191

\bibitem[Kato et~al.(2003b)]{VSNET}
  Kato, T., Uemura, M., Ishioka, R., Nogami, D., Kunjaya, C., Baba, H., \&
  Yamaoka, H.\ 2003b, \pasj, \inpress\astroph{0310209}

\bibitem[Kato et~al.(2002a)]{kat02cccnc}
  Kato, T., Uemura, M., Ishioka, R., \& Pietz, J.\ 2002a, \pasj, 54, 1017

\bibitem[Kato et~al.(2002b)]{kat02v592her}
  Kato, T., Uemura, M., Matsumoto, K., Kinnunen, T., Garradd, G., Masi, G., \&
  Yamaoka, H.\ 2002b, \pasj, 54, 999

\bibitem[Lasota et~al.(1995)]{las95wzsge}
  Lasota, J.~P., Hameury, J.~M., \& Hur\'{e}, J.~M.\ 1995, \aap, 302, L29

\bibitem[Littlefair et~al.(2003)]{lit03CVBD}
  Littlefair, S.~P., Dhillon, V.~S., \& Martin, E.~L.\ 2003, \mnras, 340, 264

\bibitem[Mennickent, Diaz(2002)]{men02CVBD}
  Mennickent, R.~E., \& Diaz, M.~P.\ 2002, \mnras, 336, 767

\bibitem[Monet et~al.(2003)]{USNOB10}
  Monet, D.~G., {et~al.}\ 2003, \aj, 125, 984

\bibitem[Nogami et~al.(1997a)]{nog97alcom}
  Nogami, D., Kato, T., Baba, H., Matsumoto, K., Arimoto, J., Tanabe, K., \&
  Ishikawa, K.\ 1997a, \apj, 490, 840

\bibitem[Nogami et~al.(1997b)]{nog97sxlmi}
  Nogami, D., Masuda, S., \& Kato, T.\ 1997b, \pasp, 109, 1114

\bibitem[O'Donoghue et~al.(1991)]{odo91wzsge}
  O'Donoghue, D., Chen, A., Marang, F., Mittaz, J. P.~D., Winkler, H., \&
  Warner, B.\ 1991, \mnras, 250, 363

\bibitem[Ohtani et~al.(1992)]{Ouda}
  Ohtani, H., {et~al.}\ 1992, Memoirs of the Faculty of Science, Kyoto
  University, Series A of Physics, Astrophysics, Geophysics and Chemistry, 38,
  167

\bibitem[Osaki(1995)]{osa95wzsge}
  Osaki, Y.\ 1995, \pasj, 47, 47

\bibitem[Osaki(1996)]{osa96review}
  Osaki, Y.\ 1996, \pasp, 108, 39

\bibitem[Osaki(1998)]{osa98suumareviewwzsge}
  Osaki, Y.\ 1998, in \ASPConf{137}{Wild Stars in the Old West}, ed. S. Howell,
  E. Kuulkers, \& C. Woodward (\PublisherASP), ~334

\bibitem[Patterson(1998)]{pat98evolution}
  Patterson, J.\ 1998, \pasp, 110, 1132

\bibitem[Patterson(2001)]{pat01SH}
  Patterson, J.\ 2001, \pasp, 113, 736

\bibitem[Patterson et~al.(1996)]{pat96alcom}
  Patterson, J., Augusteijn, T., Harvey, D.~A., Skillman, D.~R., Abbott, T.
  M.~C., \& Thorstensen, J.\ 1996, \pasp, 108, 748

\bibitem[Patterson et~al.(1981)]{pat81wzsge}
  Patterson, J., McGraw, J.~T., Coleman, L., \& Africano, J.~L.\ 1981, \apj,
  248, 1067

\bibitem[Patterson et~al.(2003)]{pat03suumas}
  Patterson, J., {et~al.}\ 2003, \pasp, \inpress\astroph{0309100}

\bibitem[Podsiadlowski et~al.(2003)]{pod03amcvn}
  Podsiadlowski, Ph., Han, Z., \& Rappaport, S.\ 2003, \mnras, 340, 1214

\bibitem[Politano et~al.(1998)]{pol98TOAD}
  Politano, M., Howell, S.~B., \& Rappaport, S.\ 1998, in \ASPConf{137}{Wild
  Stars in the Old West}, ed. S. Howell, E. Kuulkers, \& C. Woodward
  (\PublisherASP), ~207

\bibitem[Skillman et~al.(2002)]{ski02j2329}
  Skillman, D.~R., {et~al.}\ 2002, \pasp, 114, 630

\bibitem[Solheim(1995)]{sol95amcvnreview}
  Solheim, J.-E.\ 1995, \BaltA, 4, 363

\bibitem[Stellingwerf(1978)]{PDM}
  Stellingwerf, R.~F.\ 1978, \apj, 224, 953

\bibitem[Stolz, Schoembs(1984)]{StolzSchoembs}
  Stolz, B., \& Schoembs, R.\ 1984, \aap, 132, 187

\bibitem[Thorstensen et~al.(2002a)]{tho02qzser}
  Thorstensen, J.~R., Fenton, W.~H., Patterson, J.~O., Kemp, J., Halpern, J.,
  \& Baraffe, I.\ 2002a, \pasp, 114, 1117

\bibitem[Thorstensen et~al.(2002b)]{tho02j2329}
  Thorstensen, J.~R., Fenton, W.~H., Patterson, J.~O., Kemp, J., Krajci, T., \&
  Baraffe, I.\ 2002b, \apjl, 567, L49

\bibitem[Thorstensen et~al.(2002c)]{tho02gwlibv844herdiuma}
  Thorstensen, J.~R., Patterson, J.~O., Kemp, J., \& Vennes, S.\ 2002c, \pasp,
  114, 1108

\bibitem[Uemura et~al.(2002)]{uem02j2329letter}
  Uemura, M., {et~al.}\ 2002, \pasj, 54, L15

\bibitem[Vogt(1980)]{vog80suumastars}
  Vogt, N.\ 1980, \aap, 88, 66

\bibitem[Warner(1985)]{war85suuma}
  Warner, B.\ 1985, in Interacting Binaries, ed. P.~P. Eggelton, \& J.~E.
  Pringle (\PublisherReidel), ~367

\bibitem[Warner(1995a)]{war95amcvn}
  Warner, B.\ 1995a, \apss, 225, 249

\bibitem[Warner(1995b)]{war95suuma}
  Warner, B.\ 1995b, \apss, 226, 187

\bibitem[Wild(1979)]{wil79lland}
  Wild, P.\ 1979, \iaucirc, 3412

\end{thebibliography}
\end{document}